\documentclass[aps,physrev,reprint,,superscriptaddress]{revtex4-2}

\usepackage{url}
\usepackage{braket}
\usepackage{amsmath}
\usepackage{tikz}
\usepackage{bbold}
\usetikzlibrary{quantikz2}
\usepackage{siunitx}
\usepackage{float}
\usepackage{booktabs} 
\usepackage{listings}

\usepackage{xspace} 

\usepackage[colorlinks]{hyperref}
\hypersetup{%
        plainpages=true,
        breaklinks=true,
        hypertexnames=false,
        pageanchor=true,
        colorlinks=true,
        linkcolor={blue},
        citecolor={blue},
        urlcolor={blue},
        anchorcolor={black}
}

\newcommand{\subfigref}[2]{\hyperref[#1]{\ref*{#1}(#2)}}

\usepackage{mleftright} 

\newcommand{\figref}[1]{\mbox{Fig.~\ref{#1}}}
\newcommand{\tabref}[1]{\mbox{Table~\ref{#1}}}
\newcommand{\secref}[1]{\mbox{Sec.~\ref{#1}}}

\newcommand{\appref}[1]{\mbox{Appendix~\ref{#1}}}
\renewcommand{\eqref}[1]{\mbox{Eq.~(\ref{#1})}}

\newcommand{\figpanel}[2]{Fig.~\hyperref[#1]{\ref*{#1}(#2)}}
\newcommand{\figpanels}[3]{Fig.~\hyperref[#1]{\ref*{#1}(#2)-(#3)}}
\newcommand{\figpanelNoPrefix}[2]{\hyperref[#1]{\ref*{#1}(#2)}}

\newcommand{\switchgate}{\texttt{Sw}\xspace}
\newcommand{\switchdelay}{\texttt{SDel}\xspace}
\newcommand{\rhototal}{\rho_\text{total}}
\newcommand{\singlequbitgatetime}{t_\text{1q}}

\begin{document}

\title{Overhead in Quantum Circuits with Time-Multiplexed Qubit Control}

\author{Marvin Richter}
\email{marvinr@chalmers.se}

\author{Ingrid Strandberg}

\author{Simone Gasparinetti}

\author{Anton Frisk Kockum}
\email{anton.frisk.kockum@chalmers.se}

\affiliation{Department of Microtechnology and Nanoscience, Chalmers University of Technology, 412 96
Gothenburg, Sweden}

\begin{abstract}

When scaling up quantum processors in a cryogenic environment, it is desirable to limit the number of qubit drive lines going into the cryostat, since fewer lines makes cooling of the system more manageable and the need for complicated electronics setups is reduced. However, although time multiplexing of qubit control enables using just a few drive lines to steer many qubits, it comes with a trade-off: fewer drive lines means fewer qubits can be controlled in parallel, which leads to an overhead in the execution time for quantum algorithms. In this article, we quantify this trade-off through numerical and analytical investigations. For standard quantum processor layouts and typical gate times, we show that the trade-off is favorable for many common quantum algorithms --- the number of drive lines can be significantly reduced without introducing much overhead. Specifically, we show that couplers for two-qubit gates can be grouped on common drive lines without any overhead up to a limit set by the connectivity of the qubits. For single-qubit gates, we find that the serialization overhead generally scales only logarithmically in the number of qubits sharing a drive line. These results are promising for continued progress towards large-scale quantum computers. 

\end{abstract}

\maketitle


\section{Introduction}
\label{sec:intro}

Solid-state quantum processors, e.g., those using superconducting qubits~\cite{Gu2017, Krantz2019Jun, Blais2021} or semiconductor quantum dots~\cite{Chatterjee2021Mar, Burkard2023Jun}, require operation at millikelvin temperatures to facilitate the systems to be initialized in their ground state and prevent thermal excitations.
In current standard architectures, qubit control and readout are performed using microwave pulses generated at room temperature. These pulsed signals are delivered via coaxial lines that connect the quantum processor to the room-temperature electronics. In order to protect the qubits from thermal radiation flowing from the higher temperature stages to lower temperature stages, the coaxial lines are filtered and thermally anchored at several stages within the cryostat~\cite{Krinner2019}. 

As quantum processors are being scaled up to larger numbers of qubits, one of several challenges that arise is the cooling~\cite{mohseni2025buildquantumsupercomputerscaling}. The cooling challenges stem from the fact that every qubit must be individually addressable. In the current small- to medium-scale systems, this addressability is typically achieved by individual connections, with each qubit having a dedicated control line. While this method is straightforward, it has two major limitations that prohibit scaling up. First, as the number of qubits increases, each line adds a thermal load on the cryostat, which has a limited cooling power. Second, there is a limit on the number of coaxial lines that can physically fit in the cryostat. 
There are various suggestions for avoiding these bottlenecks, such as using photonic links instead of coaxial cables~\cite{Lecocq2021Mar, Youssefi2021May, Weaver2024Feb, Shen2024Apr} or using single-flux-quantum logic~\cite{McDermott2018Jan, Liu2023Jul, Mukhanov2019}. Another option is to perform frequency- or time-domain multiplexing via a controller mounted on the cold stage of the cryostat, close to the quantum chip~\cite{Huang2022Dec, hornibrook2015, vanDijk2020Sep, mohseni2025buildquantumsupercomputerscaling, Zhao2024Mar}.

While frequency-multiplexed readout is well established~\cite{Chen2012Nov, Heinsoo2018Sep}, frequency-multiplexed control is still in its early stages~\cite{Asaad2016Aug, Yang2024May, Ohira2024Sep, Matsuda2025Jan}. In contrast, time-multiplexed control has received little direct attention, even though enabling hardware such as CMOS-based switches have been demonstrated to enable efficient characterization of large numbers of semiconductor quantum dots~\cite{Ward2013May, Al-Taie2013Jun, PaqueletWuetz2020May, Thomas2025Jan}, and are also considered for the purpose of scalable quantum computing~\cite{Gonzalez-Zalba2021Dec, Bohuslavskyi2024Oct, Bavdaz2022Jul, Bartee2025Jul}.
A CMOS device for time-multiplexed superconducting qubit control, working at 10 mK, has also been demonstrated~\cite{Acharya2023}. Additionally, there are significant efforts to create superconducting switches~\cite{Pechal2016Aug, Chapman2016May,Naaman2016Mar, DeSimoni2018Sep, Wagner2019Oct,Graninger2023Apr, Zotova2024Feb, Huang2024May} and semiconducting-superconducting hybrid devices~\cite{Paghi2024Oct, Paghi2025May, Barati2021Mar} to reduce heat dissipation compared to CMOS technology.
While such devices are under active development, a systematic study of their prospected impact in quantum computing systems is missing. Here, we analyze the feasibility of time multiplexing approaches for qubit control.

\begin{figure*}
    \centering
    \includegraphics[width=\linewidth]{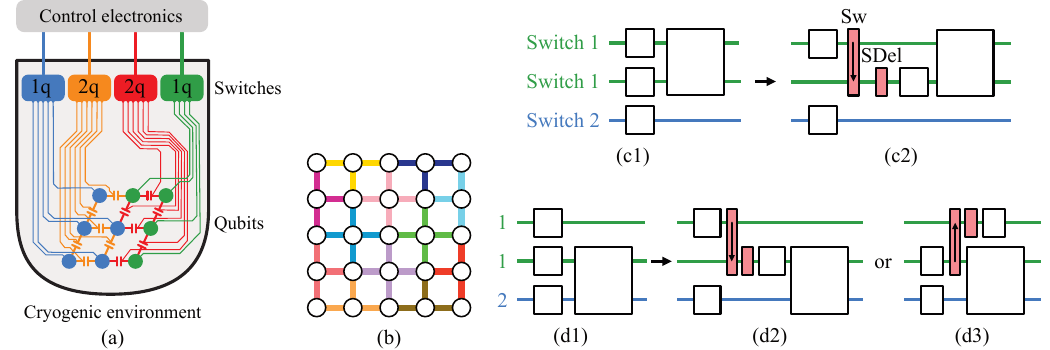}
    \caption{Time-multiplexed qubit control and compilation considerations.
    (a) A schematic drawing of control-signal demultiplexing via switches to qubits and coupling elements. 
    (b) Layout for a 5$\times$5 square-grid quantum processor with twelve color-coded coupling control groups for two-qubit gates, which minimizes required control lines without increasing the duration of any quantum circuit. 
    (c) Circuit serialization modeling for single-qubit gates: (c1) target circuit and (c2) serialized circuit with controller switch gates \switchgate (red arrow) and delay gates \switchdelay (red) representing switching time. Quadratic blocks represent single- and two-qubit gates.
    (d) Impact of gate-execution ordering on overhead: (d1) target circuit, (d2) default ordering by qubit index, and (d3) distance-optimized ordering based on subsequent two-qubit gates.}
    \label{fig:combined-fridge-grid25-circuits}
\end{figure*}

In this article, we numerically and analytically study the impact on quantum computing performance when a time multiplexing approach is adopted to facilitate scaling quantum processors. The approach we consider is one in which the number of lines are reduced by controlling multiple qubits sequentially via a single drive line~\footnote{Technically, this describes demultiplexing, but we simply refer to it as multiplexing for convenience.}; see \figpanel{fig:combined-fridge-grid25-circuits}{a}. 
If multiple qubits with different frequencies share a single drive line, only one qubit can be driven at a time, meaning the other qubits must idle during that period. Furthermore, the switching itself will introduce some idling time for all qubits. This sequential control thus introduces an overhead in the runtime of a given algorithm. Given the limited qubit coherence time, it is natural to ask how large this overhead is and whether it has a negative impact on the fidelity of quantum algorithms~\cite{Abad2022}. 

We develop a compilation algorithm~\cite{Ge2024} to minimize the runtime overhead from time multiplexing. Although a few compilation methods for similar purposes have been put forward~\cite{Shan2022May, Lao2021Feb, Huang2024, Booth2018Jun}, they have only been benchmarked on relatively small processors with a focus on the compiler performance itself. Here, using the open-source quantum software stack Qiskit~\cite{qiskit} to serialize qubit operations and emulate idling times due to multiplexer hardware constraints, we investigate the resulting runtime overhead when using our compilation algorithm for random circuits and a benchmark set of quantum algorithms~\cite{Quetschlich2023mqtbench} on different quantum processor layouts, and focus on the consequences of this overhead for multiplexing design. In particular, the processor layouts we consider are a square-grid layout similar to the Google Sycamore and Willow processors~\cite{Arute2019Oct, Acharya2025} with up to 121 qubits, and the heavy-hexagon layout of the 127-qubit IBM Eagle-type processor~\cite{eagle_blogpost}.

We find that time multiplexing enables reducing the number of control lines in quantum computers while only paying a low price in circuit runtime overhead. The number of lines controlling couplers for two-qubit gates can even be reduced without overhead by a factor determined by the qubit connectivity of the quantum processor layout, since a single qubit is connected to multiple couplers but only can take part in one two-qubit gate at a time. For single-qubit gates, we observe that reducing the number of control lines by a factor $k$ in most cases only generates an overhead scaling logarithmically with $k$, and we are able to explain this finding using queueing theory. We also make several more observations about the runtime for implementations of specific standard quantum algorithms. Overall, our main findings are encouraging for the development of larger quantum computers --- the need for fewer control lines does not appear likely to become a limiting factor in that endeavor.

This article is organized as follows. In \secref{sec:time-multiplexing}, we present the considered setup, discuss the impact of time multiplexing on two-qubit gates, describe how we model and compile circuits with time multiplexing of single-qubit control, and detail how we benchmark our methods. The extent of the serialization overhead from time multiplexing of single-qubit control depends on many factors, such as the circuit to be executed, gate durations, and the number of qubits sharing a switch. We explore variations of these parameters in Sec.~\ref{sec:results} and make observations about how their impact on the overhead scales. Finally, we summarize our findings and present our conclusions in \secref{sec:summary_conclusion}, and discuss avenues for future research in \secref{sec:outlook}.


\section{Time-multiplexed qubit control}
\label{sec:time-multiplexing}

Here, we describe our approach to compilation for time-multiplexed qubit control in more detail and explain how we benchmark this approach. First, we present general considerations and a discussion of related work in \secref{sec:general-related}. Then, we discuss the impact of time multiplexing on two-qubit gates in \secref{sec:twoqubit-gates} and show that these permit a limited degree of multiplexing without overhead, and we therefore focus our study on the impact of single-qubit multiplexing.
For single-qubit gates, we explain and motivate our modeling and compilation approach in \secref{sec:modeling-serialization-1q}. Finally, we describe, in \secref{sec:benchmarking}, how we benchmark the compilation of time-multiplexed control for single-qubit gates.


\subsection{General considerations and related work}
\label{sec:general-related}

To evaluate the feasibility of time-multiplexed qubit control, we consider a setup with superconducting qubits. A schematic overview of a representative multiplexing setup is shown in \figpanel{fig:combined-fridge-grid25-circuits}{a}. Here, control electronics outside the dilution refrigerator connect to (de)multiplexers or switches at the cold stage of the cryostat, which in turn route control signals to selected qubits. Different colors represent different switches and the corresponding connected qubits or coupling elements. 
Drive signals for single-qubit gates on qubits are separated from drives to couplers that mediate two-qubit gates between qubits.
This structure is common~\cite{Chen2014Nov, McKay2016Dec, Yan2018Nov} and used in several high-profile experiments~\cite{Arute2019Oct, Acharya2023_google, Acharya2025}.

The presence of time-multiplexed control introduces a new step in the procedure for mapping a logical quantum circuit onto one physically executable on the quantum processor.
This procedure is generally referred to as quantum circuit compilation~\cite{Kusyk2021Mar, Ge2024}. Two critical steps in quantum circuit compilation without time-multiplexed control are translation into native gates of the hardware in question, and qubit routing, which inserts \texttt{SWAP} gates to perform arbitrary two-qubit gates on a layout with limited connectivity~\cite{Li2019, Cowtan2019, Nannicini2022, Ito2023}.   
To emulate the constraint of sequential qubit operations imposed by routing control signals through shared switches, we add a third step: serialization of concurrent gates that share the same switch. When we speak of serialization overhead in the rest of article, we include both the time added from the serial rather than parallel application of gates and the time delay due to switching between the different control lines to execute the gates in series.

A few related works have previously explored compilation in such settings, though our article differs from these in key aspects.
For instance, Ref.~\cite{Shan2022May} presented a compilation method designed for multiplexed quantum control, where the main benchmark was utilization efficiency of the classical control channels rather than quantum circuit or algorithm performance.
References~\cite{Lao2021Feb, Huang2024} introduced compilers that include constraints imposed by shared classical control electronics, and a constraint functionally similar to shared controls was imposed in~\cite{Booth2018Jun}, where neighboring qubits were restricted from simultaneous operation due to crosstalk, similarly as if they shared a control line.
These methods were tested on between 3 and 21 qubits, while our work here includes benchmarks on up to 127 qubits. Most importantly, these previous works focused entirely on the performance of the compilers, while our study considers the runtime of the compiled circuits and what consequences that has for time-multiplexing schemes.
Hence, we do not perform a fully optimized serialization, but implement a strategic serialization procedure that avoids superfluous switching and utilizes slack time induced by long two-qubit gates. Our serialization procedure is described in~\secref{sec:modeling-serialization-1q}.


\subsection{Impact on two-qubit gates}
\label{sec:twoqubit-gates}

Time multiplexing control of single-qubit and two-qubit gates present different considerations, so we need to treat them separately. In our hardware model, which reflects common superconducting processor designs, two-qubit gates are implemented by driving couplers positioned between qubits. To alleviate the hardware overhead from individual drive lines, we propose employing the following time multiplexing strategy for coupler lines: couplers surrounding alternating qubits are grouped together, as illustrated in \figpanel{fig:combined-fridge-grid25-circuits}{b}. For the 5$\times$5 qubit array, this multiplexing approach reduces the required coupler control lines from 40 to 12.

While concurrent two-qubit gates sharing the same control switch would have to be serialized for arbitrary coupler groupings, this proposed grouping avoids scheduling conflicts by exploiting the constraint that each qubit can participate in at most one two-qubit gate per time step, ensuring that control signals within a group are naturally separated in time. 
This type of natural serialization arises, e.g., in surface-code stabilizer measurements in quantum error correction~\cite{Fowler2012Sep, Acharya2025}.

We thus see that time multiplexing control can reduce the number of control lines required for two-qubit gates essentially for free, up to a reduction factor given by the connectivity of the qubits. For a square grid, this reduction factor is four in the asymptotic limit; for a heavy-hexagon layout, it is three (grouping the couplers connected to each hexagon corner). Although one can consider larger groups of couplers, which would results in serialization overhead, we are content with the ``free'' reduction for now. Accordingly, we assume in the remainder of this article that \emph{serialization of two-qubit gates introduces no overhead} in circuit runtime. Therefore, we focus the rest of our analysis on serialization of single-qubit gates.

We note that there are other schemes for two-qubit gates, which do not rely on controlling a coupler. A prominent example is the cross-resonance gate~\cite{Paraoanu2006, Rigetti2010, Chow2011, Sheldon2016a, Heya2021}, where one qubit is directly driven at the transition frequency of another qubit. For the purposes of this work, such gates are similar to single-qubit gates, and thus partly covered by the investigation of single-qubit gates in the remainder of this article. There are also cross-resonance three-qubit gates~\cite{Kim2021, Itoko2024}, where multiple qubits are directly driven. Ensuring that such gates can be performed could be achieved by a suitable grouping of drive lines for single-qubit gates, with drive lines for neighboring qubits assigned to different switches; this type of grouping is one of the options discussed for single-qubit gates below.

We also note here that three-qubit gates can be performed by simultaneously controlling the two couplers in a three-qubit chain~\cite{Gu2021, Warren2023, Liu2025}. For such gates, the two couplers involved would need to belong to different switches. If we use the layout in \figpanel{fig:combined-fridge-grid25-circuits}{b}, reducing the number of coupler lines by a factor four, half of the qubits (those surrounded by four couplers belonging to the same switch) could not act as the middle qubit in these three-qubit gates, but could act as one of the end qubits. Similarly, the other half of the qubits could not act as one of the end qubits in such three-qubit gates, but could act as the middle qubits. The number of possible three-qubit gates is thus halved in this setup. A detailed analysis of this kind of trade-off for three-qubit gates is beyond the scope of the current work.


\subsection{Modeling serialization of single-qubit gates}
\label{sec:modeling-serialization-1q}

\newcommand{\gateswitchgate}[1]{
    \gate[#1]{\text{CSw}}
}
\newcommand{\gateswitchdelay}{
    \gate{\text{Cdel}}
}

For single-qubit gates, time multiplexing of control signals will generally introduce an overhead in the circuit duration. This can be easily understood, as only one qubit per switch can be controlled at a time. Effectively, concurrent single-qubit gates applied to qubits of the same switch must be serialized over time. 

Numerical analysis of the overhead caused by time-multiplexed single-qubit control requires a circuit representation model of the switch. Therefore, we introduce two dummy gates: a single-qubit switch delay \switchdelay, corresponding to the duration of a switching event, and a two-qubit switch gate \switchgate used for bookkeeping active qubits, each acting logically as the identity. While the two-qubit switch gate forces serialization via dependency upon single-qubit gates of the same qubit control group, the switch delay gate only acts on the qubit awaiting the drive signal. 

We illustrate circuit serialization with these dummy gates in a simple example in~\figpanel{fig:combined-fridge-grid25-circuits}{c}. First,~\figpanel{fig:combined-fridge-grid25-circuits}{c1} shows a circuit with parallel single-qubit gates, where two of the qubits share a switch. The resulting serialized circuit is shown in~\figpanel{fig:combined-fridge-grid25-circuits}{c2}: after a single-qubit gate is executed on the first qubit, a \switchgate gate is applied between the first and second qubit and then a \switchdelay gate is applied on the second qubit to account for nonzero switching time before the actual single-qubit gate is applied on the second qubit. 

We determine circuit execution time by calculating the critical path, i.e., the longest sequence of operations from qubit initialization to final measurement, in the circuit's directed acyclic graph (DAG) representation~\cite{childs_et_al:LIPIcs.TQC.2019.3}, where individual gate durations serve as edge weights. The zero-duration \switchgate gate does not directly contribute to the critical path length; however, it enforces temporal dependencies between previously concurrent single-qubit operations within the same control group, potentially extending the overall execution time. In contrast, the \switchdelay gate, applied to the target qubit during control transitions, introduces a brief delay $t_\text{sw}$ that may directly lengthen the critical path by increasing the cumulative duration of the longest operational sequence.

During the serialization stage, it matters in which order the single-qubit gates are processed. Since two-qubit gates are typically longer than single-qubit gates, single-qubit gates should be prioritized in the order of the smallest distance to the next two-qubit gate applied to the gate's qubit. By applying this heuristic, we bias toward situations like that depicted in \figpanel{fig:combined-fridge-grid25-circuits}{d}, where switching occurs during a concurrent two-qubit gate and thereby does not increase the total duration. We provide more details on our optimization strategy for scheduling single-qubit gates in \appref{sec:serialization-optimization}.

We employ balanced group sizes where qubits are partitioned into groups containing either $k$ or $k-1$ qubits, where $k$ represents the number of qubits per switch. This approach maximizes switch-capacity utilization, avoiding highly uneven group distributions that would lead to inefficient hardware resource allocation.

We tested different grouping strategies to assign qubits to switches. The ``trivial'' layout creates assignments based on qubit indices. Since nearby qubits typically have close indices, the trivial layout resembles the ``clustering'' layout based on qubit proximity. We also considered other layout strategies, but due to comparatively good performance (low overhead) and ease of reproducibility, we use the trivial layout for all following discussions in the main text. We refer the interested reader to \appref{app:grouping} for further details and performance of other qubit grouping strategies.


\subsection{Benchmarking}
\label{sec:benchmarking}

Here, we describe the target hardware and circuits used to quantify the overhead introduced by time multiplexing.
This overhead depends on a variety of circuit and hardware parameters, but we identify and focus on a few dominant contributors. The number of qubits per switch,~$k$,~is one such principal factor for the overhead. Since single-qubit gates are serialized, the number of single-qubit gates $N_1$ in a circuit is a relevant factor as well.
We also define and consider the gate densities of single- and two-qubit gates in a quantum circuit,
\begin{equation}
    \rho_1 = \frac{N_1}{nD}, \quad \rho_2 = \frac{2N_2}{nD}, \quad \rhototal = \rho_1 + \rho_2,
    \label{eq:gate-densities}
\end{equation}
with $n$ the number of qubits, $N_2$ the number of two-qubit gates, $D$ the circuit depth, and $\rhototal$ the total gate density.


\subsubsection{Hardware platforms}
\label{sec:benchmarking-hardware}

The exact duration overhead depends on several parameters of the target hardware. First, the native gate set determines the number of gates in the translated circuit. Second, the connectivity influences the single- and two-qubit gate densities of the executable circuit through routing. Third, the gate durations naturally affect overall execution time.
To cover a range of noisy intermediate-scale quantum (NISQ)~\cite{Preskill2018} devices, we begin with a relatively small 5$\times$5 square grid of 25 qubits, and then consider two larger systems of similar size, an 11$\times$11 square grid with 121 qubits and an IBM Eagle device with 127 qubits in a heavy-hexagon layout.

For the square-grid architecture, we divide the native gates into three groups and assume their durations as follows: i) two-qubit gates with a duration of \SI{200}{\nano\second}, ii) single-qubit gates with a duration of \SI{20}{\nano\second}, and iii) the virtual \texttt{RZ} gate that can be merged into adjacent gates as a phase shift and takes zero time to execute~\cite{McKay2017Aug}. The first group includes two-qubit gates \{\texttt{CZ}, \texttt{iSWAP}\} that can both be executed with a simple flux-tunable coupler configuration~\cite{McKay2016Dec, Krizan2025}. The second group consists of single-qubit gates \{\texttt{RX}, \texttt{RY}, \texttt{H}\} with finite duration, implemented via microwave pulses on the qubit.

These choices are motivated by experimental developments. Google devices use single-qubit gates with durations between \SI{18}{}--\SI{25}{ns} (see the Supplementary Material of Ref.~\cite{Acharya2025}). 
While flux-driven two-qubit gates have decreased their duration from around \SI{200}{ns} in early demonstrations~\cite{McKay2016Dec, Caldwell2018Sep} down to \SI{40}{}--\SI{50}{ns} in more recent work~\cite{Li2024Nov, Ding2023Sep}, many alternative approaches remain slower.
Since flux-tunable devices are susceptible to decoherence due to flux noise, fully microwave-driven couplers have been developed. These tend to have longer gate times, with recent implementations reporting between \SI{200}{ns} and \SI{380}{ns}~\cite{Shirai2023Jun, Xu2025Mar}.
Novel cat–transmon hybrid architectures also exhibit comparatively long gate times in the range of \SI{200}{ns} to \SI{400}{ns}~\cite{Putterman2025Feb, Hann2025Jul}.
Systems with long-range couplings have recently demonstrated two-qubit gates with durations between \SI{160}{ns} and \SI{430}{ns}~\cite{Deng2025Jan, Song2025Jul}, and a recently introduced multielement coupler similarly reports a duration of about \SI{340}{ns}~\cite{Huber2025Jul}. 
A two-qubit gate duration of of \SI{200}{\nano\second} therefore represents a conservative but realistic estimate.
Differences in absolute gate durations do not change the qualitative behavior and scaling of the serialization overhead, although the absolute overhead times are affected, as further discussed in \secref{sec:scaling}. 

For IBM's Eagle platform~\cite{eagle_blogpost} with a heavy-hexagon coupling map, we use the Brisbane quantum processor. This processor has single-qubit gates \texttt{SX} and \texttt{X} with a fixed duration of \SI{60}{\nano\second}, and two-qubit gate \texttt{ECR} with a duration of \SI{660}{\nano\second}~\cite{brisbane_json}. 


\subsubsection{Benchmarking circuits}
\label{sec:benchmarking-circuits}

We use two different sets of circuits to estimate the overhead introduced by time multiplexing qubit control: a) random circuits and b) examples of quantum algorithms provided in the standard benchmarking set MQT Bench~\cite{Quetschlich2023mqtbench}. 

We generate the random circuits in native gates of the target hardware with single-qubit and two-qubit gate weights 0.7 and 0.3, respectively. At each generation step, a decision between single- or two-qubit is made according to the weights; then a corresponding gate is drawn from the native gate set and added to the back of the circuit. By this procedure, we generate sparse random circuits with a single-qubit gate density $\rho_1$ of up to 0.35 for the 25-qubit square grid and up to 0.25 for the larger architectures.
We then optimize the random circuits by first using qiskit's default transpiler pass managers on optimization level 2 for the 25-qubit device and on level 1 for the larger devices to receive the target quantum circuit.

For the square-grid architecture, we retrieved the MQT Bench circuits in a hardware-independent form and translated them; for the IBM Eagle architecture, we retrieved the circuits directly in native gates.
Afterwards, we routed each native-gate circuit according to the hardware coupling map before serialization. We performed the steps separately to measure the overhead arising from both routing and serialization. Results are reported as the median over multiple transpiler seeds, in order to average out variations due to heuristic qubit routing.

Routing is a well-known source of overhead, and we compare the runtime overheads arising from routing and serialization. Because routing is a necessary step for physically executing a circuit, we analyze both the absolute runtime overhead of serialization and its value relative to the runtime of the routed circuit.


\section{Results for impact of time multiplexing on single-qubit gates}
\label{sec:results}

In this section, we present our results on time multiplexing for single-qubit gates, using the benchmarks described in \secref{sec:benchmarking} and the compilation procedure shown in \secref{sec:modeling-serialization-1q}. As discussed in \secref{sec:benchmarking-hardware}, we ran benchmarks on both grid and heavy-hexagon qubit layouts. We focus on the results for grid layouts here and defer the results for the heavy-hexagon layout to \appref{app:heavy-hexagon}, since the two layouts yield quite similar behavior for the quantities we are interested in. Some additional results for grid layouts are shown in \appref{app:square-grid}.

We begin with results for random circuits in \secref{sec:benchmarks-random} and then continue with results for quantum algorithms from MQT Bench in \secref{sec:benchmarks-mqt}. Finally, in \secref{sec:scaling}, we discuss our observations on how the time multiplexing overhead scales as a function of different parameters.


\subsection{Random circuits}
\label{sec:benchmarks-random}

\begin{figure*}
    \centering
    \includegraphics[width=1\linewidth]{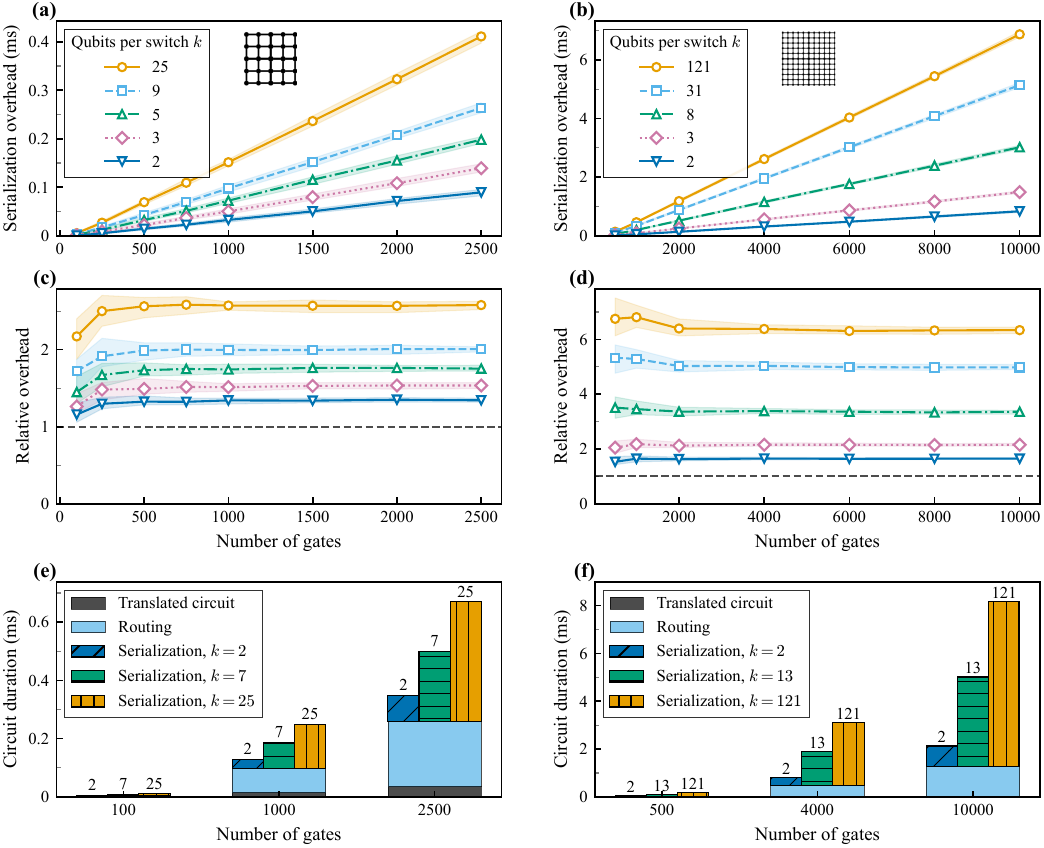}
    \caption{Overhead from time multiplexing single-qubit control for square-grid architectures (left column: 5$\times$5 grid; right column: 11$\times$11 grid). Each data point displayed is the median over 100 random circuit seeds; shaded areas show the interquartile range.  
    (a, b) Serialization overhead as an increase in circuit duration compared to the routed circuit, as a function of the number of gates in random circuits. 
    (c, d) The same serialization overhead as in (a, d), but now shown relative to the routed circuit duration, as a function of the number of gates in random circuits. 
    (e, f) Breakdown of total circuit duration: translated circuit~(black), qubit routing overhead (light blue), and serialization overhead for $k$ qubits per switch (blue, green, yellow), for random circuits with three different gate counts.
    The procedure for generating and compiling the circuits is detailed in \secref{sec:time-multiplexing}.}
    \label{fig:random-grids}
\end{figure*}

Figure~\ref{fig:random-grids} displays the overhead due to time-multiplexing of single-qubit control for random circuits, as a function of number of gates in the circuit. The panels in the left column of the figure [\figpanel{fig:random-grids}{a}, \figpanelNoPrefix{fig:random-grids}{c}, and~\figpanelNoPrefix{fig:random-grids}{e}] show results for a 5$\times$5 square grid, while the panels in the right column [\figpanel{fig:random-grids}{b}, \figpanelNoPrefix{fig:random-grids}{d}, and \figpanelNoPrefix{fig:random-grids}{f}] show results for an 11$\times$11 square grid. We provide similar results for a heavy-hexagon layout with 127 qubits in \figref{fig:line-random-eagle} in \appref{app:heavy-hexagon}.

In the top row of \figref{fig:random-grids} [\figpanel{fig:random-grids}{a} and \figpanel{fig:random-grids}{b}], we plot the overhead, in milliseconds, due to serialization of single-qubit gates, for different numbers of qubits per switch. For all cases, we observe that this overhead increases linearly with the number of gates in the circuits, with a steeper slope for more qubits per switch. These behaviors are expected since the number of gates requiring serialization is proportional to the total number of gates, and switches controlling more qubits require more serialization. We return in \secref{sec:scaling} to the question of how exactly the overhead scales with the number of qubits per switch.

In the middle row of \figref{fig:random-grids} [\figpanel{fig:random-grids}{c} and \figpanel{fig:random-grids}{d}], we show the ratio between the duration of the serialized circuits and the duration of the routed circuits, as a function of the number of gates in the circuit. Again, we find a pattern that holds in all cases: with the exception of some minor variations for a small number of gates in the circuit, this ratio remains constant when the number of gates in the circuit grows. This pattern is consistent with the linear growth of the serialization overhead observed in the top row of \figref{fig:random-grids}, since we also expect the depth of the routed circuit to grow approximately linearly with the number of gates in the original circuit (for a fixed gate density).
The value of the overhead ratio scales with the size of the processor, with more overhead for the larger processor. For example, reducing the number of lines by half gives an overhead factor 1.3 for the 5$\times$5 grid and 1.6 for the 11$\times$11.

\begin{figure*}
    \centering
    \includegraphics[width=1\linewidth]{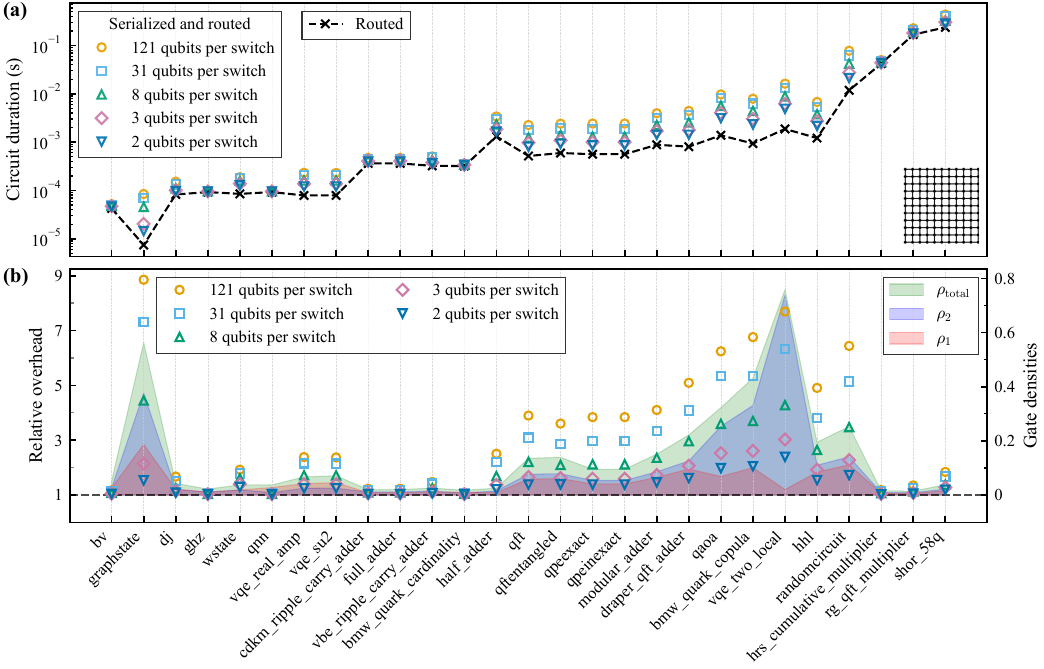}
    \caption{Impact of time multiplexing on execution of quantum algorithms from the MQT Bench set for an 11$\times$11 grid. 
    (a)~Total circuit duration for serialized routed circuits (colored markers) and routed circuits without serialization (black crosses and dashed line) across different quantum algorithms, ordered by increasing gate count. 
    (b)~Relative overhead of serialization compared to the routed circuit duration (markers, left axis), with shaded areas indicating gate densities ($\rho_1$ pink, $\rho_2$ blue, $\rho_{\rm total}$ green; right axis) and different marker colors indicating different numbers of qubits per switch. The data points are medians over 20 transpiler seeds. The full names of the quantum algorithms are listed in \tabref{tab:fit_parameters} in \appref{app:modeling_scaling}.}
    \label{fig:line-mqt-grid121}
\end{figure*}

In the bottom row of \figref{fig:random-grids} [\figpanel{fig:random-grids}{e} and \figpanel{fig:random-grids}{f}], we display examples of the full breakdown of the contributions to the total circuit duration from routing and serialization, for selected numbers of gates and qubits per switch. From these plots, we see that the routing results in a substantial overhead in circuit duration compared to the translated circuit, which helps put the overhead from serialization into perspective. For both the 5$\times$5 and 11$\times$11 square grids, having two qubits per switch results in less serialization overhead than routing overhead. The number of qubits per switch that result in serialization overhead equal to the routing overhead differs for the two grid sizes: for the 5$\times$5 grid, we see from \figpanel{fig:random-grids}{c} that it is around nine; for the 11$\times$11 grid, we see from \figpanel{fig:random-grids}{d} that it is around three. We attribute this difference to the scaling for routing overhead with grid size.


\subsection{Quantum algorithms}
\label{sec:benchmarks-mqt}

In \figref{fig:line-mqt-grid121}, we show results for time multiplexing overhead for quantum algorithms in the MQT Bench set, on an 11$\times$11 square grid. We provide similar results for a 5$\times$5 square grid in \figref{fig:line-mqt-grid25} in \appref{app:square-grid} and for a heavy-hexagon layout with 127 qubits in \figref{fig:line-mqt-eagle} in \appref{app:heavy-hexagon}.

In \figpanel{fig:line-mqt-grid121}{a}, we plot the total circuit duration when using different numbers of qubits per switch (colored markers) and compare it to the duration of the routed circuit (black crosses), for the different algorithms in the MQT Bench set (ordered according to the number of gates in the original circuits, with the fewest gates on the left). The total circuit duration and absolute multiplexing overhead can be compared to qubit coherence times to give insights into the effect of multiplexing on algorithmic fidelity~\cite{Abad2022}. We observe from the plot that the serialization overhead differs a lot from algorithm to algorithm. 
For a zoomed-in view of the cases with 2 and 4 qubits per switch, see \figpanel{fig:mqt-detail-grids}{a} in \appref{app:square-grid}; the corresponding plots for a 5$\times$5 square grid and for a heavy-hexagon layout with 127 qubits are given in \figpanel{fig:mqt-detail-grids}{b} in \appref{app:square-grid} and
in \figref{fig:mqt-detail-eagle} in \appref{app:heavy-hexagon}, respectively.

To better understand how different algorithms give rise to different serialization overhead, we plot the data from \figpanel{fig:line-mqt-grid121}{a} in terms of the relative overhead factor, defined as the ratio of serialized (and routed) circuit duration to routed circuit duration in \figpanel{fig:line-mqt-grid121}{b}, and display, for each quantum algorithm, the single- and two-qubit gate densities $\rho_1$ and $\rho_2$ along with the total gate density $\rho_{\rm total}$ [see \eqref{eq:gate-densities}]. Here, a pattern emerges: higher gate densities generally result in higher serialization overhead. This is expected since sparse circuits (with low gate density) have little need for serialization.

\begin{figure}
    \centering
    \includegraphics[width=1\linewidth]{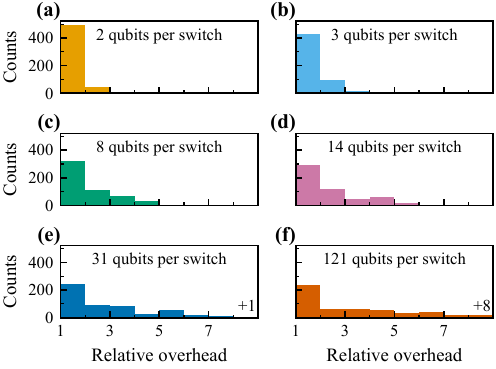}
    \caption{Distribution of time multiplexing overhead across switch sizes on an 11$\times$11 grid, for the quantum algorithms from MQT Bench shown in \figref{fig:line-mqt-grid121}, using 20 different transpilation seeds. The histograms show the relative overhead (ratio of serialized to routed circuit duration) for different numbers of qubits per switch. The $+n$ notation to the right in the bottom histograms indicates $n$ additional counts beyond the histogram limits.}
    \label{fig:hist-mqt-grid121}
\end{figure}
In \figref{fig:hist-mqt-grid121}, we provide a different perspective on the data in \figpanel{fig:line-mqt-grid121}{b}. Here, we display the distribution of serialization overhead relative to the routed circuit duration, for six different numbers of qubits per switch, ranging from 2 to 121. As the number of qubits per switch increases, large serialization overheads become increasingly common. We provide the corresponding results for a 5$\times$5 square grid in \figref{fig:hist-mqt-grid25} in \appref{app:square-grid} and for a heavy-hexagon layout with 127 qubits in \figref{fig:hist-mqt-eagle} in \appref{app:heavy-hexagon}.

\subsection{Scaling behavior}
\label{sec:scaling}

\begin{figure}
    \centering
    \includegraphics[width=1\linewidth]{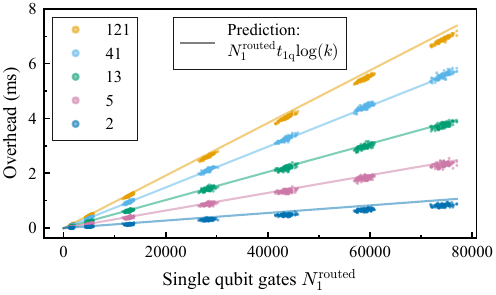}
    \caption{Scaling of serialization overhead for random circuits on an 11$\times$11 square grid. For each gate count, 100 random circuits were generated. The number of single-qubit gates~$N_1^\text{routed}$ is retrieved after routing, with slight variations due to optimizations and routing. The serialization overhead scales linearly with $N_1^\text{routed}$ and the single-qubit gate duration~$t_\text{1q}$, but logarithmically with qubits per switch $k$. Predictions (solid lines) align well with the data without additional scaling factors.}
    \label{fig:prediction-random}
\end{figure}

\begin{figure}
    \centering
    \includegraphics[width=1\linewidth]{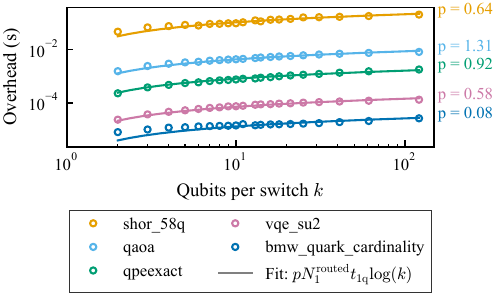}
    \caption{Scaling of serialization overhead for selected quantum algorithms from the MQT Bench set on an 11$\times$11 grid. The overhead is fitted using an additional (compared to \figref{fig:prediction-random}) scaling factor $p$ (values shown next to each fit line) and scales logarithmically with qubits per switch $k$. The complete data for all MQT Bench algorithms is shown in \figref{fig:fits-mqt-all} in \appref{app:modeling_scaling}.}
    \label{fig:prediction-mqt}
\end{figure}

We further explore the functional dependence of the serialization overhead on the dominant parameters. As could be expected, we observe that the overhead increases linearly with the number of single-qubit gates~$N_1$ in the routed circuit, and the single-qubit gate time $\singlequbitgatetime$. We saw this behavior in \figpanel{fig:random-grids}{a} and \figpanel{fig:random-grids}{b} for~5$\times$5 and 11$\times$11 grids, respectively. We further show that this linear dependence holds for the~11$\times$11 grid with random circuits in \figref{fig:prediction-random} and with circuits from MQT Bench in \figref{fig:prediction-mqt}, where data points are compared to the predicted linear function (solid lines) for different numbers of qubits per switch.  

Naively, one might also expect a linear scaling of the serialization overhead in the number of qubits per switch~$k$, since each of the~$k$ qubits must be addressed sequentially (illustrated in Fig.~\ref{fig:dense-serialization} in \appref{app:modeling_scaling}).
However, this turns out to not hold in general. For all benchmark circuits, we observe a logarithmic scaling in~$k$, as shown by the solid lines in Figs.~\ref{fig:prediction-random} and \ref{fig:prediction-mqt}. For the MQT Bench circuits in Fig.~\ref{fig:prediction-mqt}, the solid lines are fits to the function~$p N_1 \singlequbitgatetime \log (k)$ with fitting parameter~$p$. The rest of the benchmarks in the MQT Bench set follow the same scaling, as we show in detail in \appref{app:modeling_scaling}.

\begin{figure*}
\centering
    \includegraphics[width=1\linewidth]{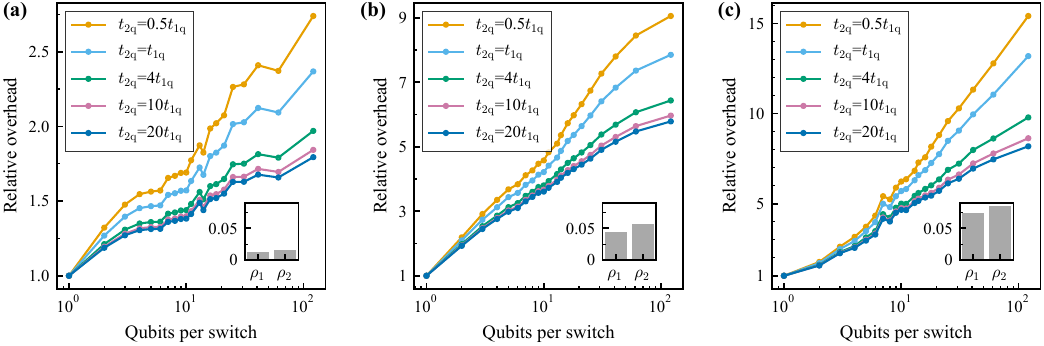}
    \caption{Relative overhead by serialization compared to the routed circuit duration, as a function of qubits per switch, for different ratios of two-qubit to single-qubit gate durations, $t_\text{2q}$ and $t_\text{1q}$, respectively, for selected algorithms on an 11$\times$11 grid. 
    (a) Shor's algorithm. 
    (b) QAOA. 
    (c) Graph-state preparation. For increasing $t_\text{2q}$, the overhead by serialization decreases as serialized single-qubit gates are more likely to be concurrent with a two-qubit gate. The insets show the single- and two-qubit gate densities of the circuit. Note the logaritmic $x$ axis. Each marker represents the median over 30 seeds.}
    \label{fig:different-t2}
\end{figure*}
    
The reduction from linear to logarithmic scaling in $k$ arises from the interplay between single- and two-qubit gates. Two-qubit gates impose a structure on the circuit, introducing idle time slots where some qubits are naturally inactive. In contrast, a circuit composed solely of single-qubit gates exhibits no dependencies, allowing all gates to be executed as soon as possible, making the circuit automatically dense. In this case, the overhead indeed scales linearly $\sim N_1 \singlequbitgatetime k$, as confirmed by benchmarks including only single-qubit gates. Besides this structural effect related to the gate density, two-qubit gates introduce another important mechanism that was mentioned in the discussion of efficient serialization in \secref{sec:modeling-serialization-1q}: during a two-qubit gate, sequences of single-qubit gates on other qubits can be performed without increasing the overall circuit duration. 
Together, these two effects can account for the observed logarithmic scaling in $k$ of the serialization overhead. A simple numerical toy model reproduces this behavior, as we show in \appref{app:modeling_scaling}, although not all parameter regimes fit the analytical reference equally well. Moreover, the dependence on two-qubit gate duration does not appear to be universal across all benchmarks, and it seems likely that the overall circuit density should be included in a complete model.

While these observations explain a sublinear scaling in~$k$, the logarithmic form is not immediately obvious. It can be rationalized using arguments from queueing theory, where extreme-value methods are applied to analyze maximum waiting times. In queueing models, the waiting time or queue length is treated as a random variable, and in many standard settings, the asymptotic behavior of the queue-length distribution has an exponentially decaying tail~\cite{Asmussen2003}. These methods are also used to analyze multiplexing and other types of signal transmission in classical communication networks~\cite{Choudhury1996Feb, Ciucu2015}. Practically, these properties of queues mean that the probability of encountering a very long queue decreases exponentially with its length, making long delays rare. In \appref{app:modeling_scaling}, we present a heuristic derivation of how this leads to logarithmic scaling.
Alternatively, in queueing-theory terms, each switch in our time multiplexing setup can be viewed as a server that sequentially processes single-qubit gates (jobs) from multiple qubits (clients). The serialization of gates on a given switch generates waiting times, and the total circuit duration is determined by the maximum waiting time across all switches. It can be calculated that the expectation value of the maximum waiting time is logarithmic in the number of clients (qubits)~\cite{Berger1995Jul}. 


\subsection{Effect of gate and switch durations}

For the simulations described above, we selected a ratio between single- and two-qubit gate durations $t_\text{2q} = 10\, t_\text{1q}$ (see \secref{sec:benchmarking-hardware}). In \figref{fig:different-t2}, we present the relative overhead for different $t_\text{2q}/t_\text{1q}$ ratios for Shor's algorithm, QAOA, and graph-state preparation. The plots show that larger $t_\text{2q}/t_\text{1q}$ ratios result in lower overhead, likely because more single-qubit gates can be executed serially during the longer two-qubit gates effectively hiding the serialization cost. The functional scaling with qubits per switch $k$ does not appear to be affected by different ratios; it remains approximately logarithmic, but with different multiplicative factors.

\begin{figure*}
\centering
    \includegraphics[width=1\linewidth]{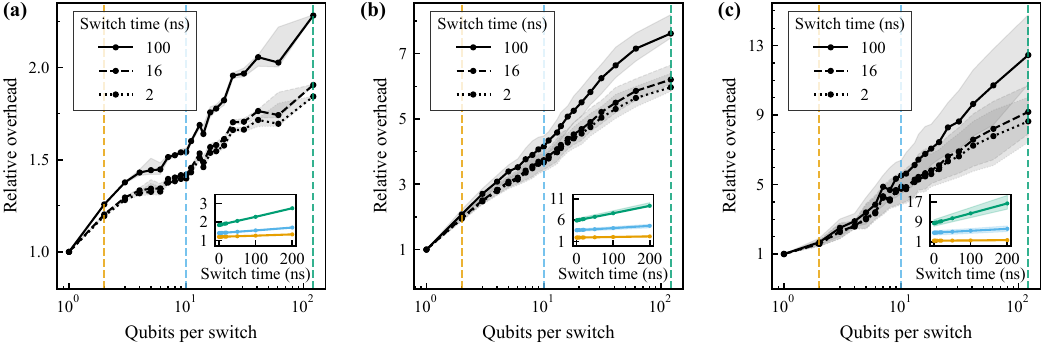}
    \caption{Relative overhead by serialization compared to the routed circuit duration, as a function of qubits per switch, for different switch delay times $t_\text{sw}$, for selected algorithms on an 11$\times$11 grid with $t_\text{2q} = 10\, t_\text{1q}$. 
    (a) Shor's algorithm. 
    (b) QAOA. 
    (c) Graph-state preparation. 
    Note the logarithmic $x$ axis. Each marker represents the median over 30 seeds and shaded areas correspond to the inter-quartile range. The insets show serialization overhead as a function of $t_\text{sw}$ for 2, 10 and 121 qubits per switch, color corresponding to the vertical dashed lines in the main plot. These plots confirms a linear dependence of the relative overhead on the switch delay time.}
    \label{fig:different-tsw}
\end{figure*}

All circuits considered so far have used a switch delay time $t_\text{sw}=\SI{2}{ns}$, corresponding to the time it takes for a switching event to occur. This is fast but realistic for state-of-the-art devices~\cite{Acharya2023}. 
The impact of longer switch times of up to \SI{100}{ns}, modeled by the \switchdelay gate in the serialized circuit, is shown in \figref{fig:different-tsw}, again for Shor's algorithm, QAOA, and graph-state preparation on the 11$\times$11 square grid. 
From \figref{fig:different-tsw}, we see that the overhead depends linearly on the switch time, with steeper increase when there are more qubits per switch. The scaling of the overhead with the number of qubits per switch remains approximately logarithmic.


\section{Summary and conclusion}
\label{sec:summary_conclusion}

We have quantified an important trade-off that occurs when scaling up quantum computers: the impact on quantum circuit duration (and hence performance) from reducing the number of control lines. For quantum computers to outperform modern supercomputers, a large number of qubits will be required, even for problems where quantum advantage is theoretically achievable. The strength of classical computing is in many ways due to the invention of the integrated circuit, which enabled the integration of millions of transistors on a small chip surface while significantly reducing the number of external connections.
Similarly, scaling up quantum computers requires more efficient control schemes than the current approach, where the number of control terminals at room temperature scale linearly with the number of qubits. To address this challenge, on-chip signal routing or multiplexing can be used to reduce wiring complexity and enable practical scaling of both superconducting- and semiconductor-based quantum computing platforms~\cite{Franke2019Jun, Borsoi2024Jan}.

The trade-off we quantified occurs since a scalable architecture with shared control lines limits the possibility of gate parallelization. This leads to increased circuit runtimes, increasing exposure to decoherence, and therefore the impact on algorithm execution needs to be assessed. 
To this end, we quantified the serialization overhead for a range of standard benchmark quantum algorithms and random quantum circuits on three NISQ devices: a 5$\times$5-qubit square grid, an 11$\times$11-qubit square grid, and IBM’s 127-qubit Eagle heavy-hexagon layout. We assumed a hierarchy of time scales common to such devices: two-qubit gates take longer to execute than single-qubit gates, which in turn are slower than the switching delay in multiplexed control. We verified that our scaling results were robust to variations of the specific times when they respected this hierarchy.

For two-qubit gates, strategic placement of controlled couplers mediating these gates on switches enables serialization of operations without incurring any duration overhead, eliminating the trade-off, up to a limit of couplers per switch set by the connectivity between qubits. For a square-grid layout, the number of control lines can be reduced by a factor of four without affecting the circuit, since each qubit is connected to four other qubits, but only can be involved in a two-qubit gate with one of them at a time. Although more couplers per switch will lead to serialization overhead, affected also by the density of two-qubit gates in the quantum circuit to be executed, we deemed the ``free'' reduction sufficient to warrant more focus on single-qubit gates. Thus, we only studied overhead from serialization of single-qubit gates in the rest of the article.

For the serialization overhead for single-qubit gates, we found a surprisingly benign, logarithmic scaling with the number $k$ of qubits per switch. Only when the quantum circuits are close to fully saturated with just single-qubit gates does the scaling become linear with $k$. We understand the logarithmic scaling with $k$ from queueing theory: the duration overhead corresponds to the maximum waiting time for all gates to be executed, and that waiting time scales logarithmically. 

Together, these findings for single- and two-qubit gates constitute the key results of our work: the number of control lines in a quantum computer can be significantly reduced without introducing much overhead in execution time for quantum algorithms. In other words, the trade-off we studied turns out to be manageable, which opens up for scaling quantum computers to large sizes. Additionally, multiplexing can be combined with moving signal generation into the cryostat, where further reduced wiring would make integration even more practical~\cite{Patra2017Sep, Pauka2021Jan, Huang2022Dec}.

We can make several further observations for specific quantum algorithms, which can be used as input for co-design of quantum processors and algorithms. Among the most well-known benchmark algorithms, GHZ state generation and quantum neural networks showed the lowest serialization overhead for single-qubit gates. 
For these algorithms, the overhead is small when using two qubits per switch, which halves the number of drive lines and provides a significant hardware reduction. Specifically, the circuit duration is extended only between \qtyrange{2}{3}{\micro\second} on the 121-qubit grid and 127-qubit heavy-hexagon architectures, while it is completely negligible for the 25-qubit grid. 

Graph-state preparation adds \SI{5}{\micro s} for the 121-qubit grid and nearly an order of magnitude more, \SI{60}{\micro\second}, for the Eagle architecture, while remaining minor on the 25-qubit grid (\SI{55}{\nano\second}). These numbers can be compared with state-of-the-art coherence times for superconducting qubits, which are in the range of hundreds of microseconds.

Several other prominent algorithms exhibit overheads that could be problematic with short coherence times. For the larger devices, QFT adds between \SI{0.2}{\milli\second} and \SI{0.9}{\milli\second}, and the additional duration for serialized QAOA (Quantum Approximate Optimization Algorithm) exceeds \SI{1}{\milli\second} on the 121-qubit grid and approaches \SI{9}{\milli\second} on Eagle. 

In general, the 127-qubit Eagle generally shows larger overhead than the 121-qubit grid, and both are higher than for the small grid. This is likely because running algorithms on more qubits requires more gates, and the overhead scales linearly with the number of gates. Eagle's overhead exceeds the 121-qubit grid on all benchmarks except W-state preparation, where both are of the order of tens of microseconds. Since Eagle only has six more qubits than the large grid, it remains unclear whether the difference is due primarily to the layout difference or the qubit count, but the lower connectivity leads to larger routing overhead for the Eagle layout.


\section{Outlook}
\label{sec:outlook}

There are several directions for future work that could be explored. One direction is to use queueing-theory methods and other analytical tools to more accurately model and optimize serialization of both single- and two-qubit gates. Further numerical studies could also help further elucidate the scaling of circuit duration overhead as a function of the number of control lines, the layout of the quantum processor, the gate densities in the quantum circuit, the switching time, the single-qubit gate time, and the two-qubit gate time. All of these factors can vary between hardware implementations, so a more detailed quantification of their impact would be valuable for choosing quantum processor architectures to scale up.

When scaling up with the goal of building a fault-tolerant quantum computer, the impact of time multiplexing on quantum error correction should be studied further using the framework we have introduced here. For a fixed number of control lines, there would be a trade-off between how large code distance one could achieve by using many qubits and how much the serialization overhead would degrade the performance of such a longer-distance code. The analysis of such a trade-off would depend heavily on details of the hardware, e.g., gate times for different types of gates, the error budget for these gates, and other error sources. For particular hardware layouts and error-correction codes, there can be highly optimized arrangements of operations in time and space~\cite{Versluis2017Sep}.

Another natural direction for future work is to integrate serialization directly into routing and compilation algorithms, enabling global optimization of both qubit placement and control-line allocation. Here, factors such as coherence times varying across a quantum processor could be taken into account. 
The setup we studied could also be combined with frequency multiplexing to further reduce the number of lines~\cite{Shi2023Dec}.

\begin{acknowledgments}
We thank David Fitzek for taking part in the initial project discussions.
We acknowledge support from the Knut and Alice Wallenberg Foundation through the Wallenberg Centre for Quantum Technology (WACQT). MR and AFK also acknowledge support from the Swedish Foundation for Strategic Research (grant number FUS21-0063). AFK is also supported by the Swedish Foundation for Strategic Research (grant number FFL21-0279) and the Horizon Europe programme HORIZON-CL4-2022-QUANTUM-01-SGA via the project 101113946 OpenSuperQPlus100.
IS acknowledges financial support from the European Union via Grant No.~101057977 SPECTRUM. SG acknowledges financial support from the European Research Council (Grant No.~101041744 ESQuAT). Numerical results were obtained using HPC resources at the Chalmers Centre for Computational Science and Engineering (C3SE). 
External interest disclosure: SG is a co-founder and equity holder in Sweden Quantum AB.
\end{acknowledgments}

\section*{Data Availability Statement}

The code and data that support the findings of this article will be openly available at \url{https://github.com/marvin-richter/overhead-time-multiplexing} upon publication.

\appendix


\section{Compilation strategies for minimizing serialization overhead for single-qubit gates}
\label{sec:serialization-optimization}

Here, we provide further details about the algorithms we considered and used for minimizing serialization overhead for single-qubit gates. We can reduce the overhead from switch delays significantly by leveraging the temporal overlap between switching operations and concurrent two-qubit gates. Since two-qubit gates have durations equivalent to multiple single-qubit operations plus switching time in most situations we consider in this article (see \secref{sec:benchmarking-hardware}), the transition of control between qubits can often be scheduled during these longer operations without extending the total circuit duration.
Our serialization model always places a zero-duration two-qubit switch gate between the source qubit and the target qubit to enforce temporal dependencies. When an actual two-qubit gate already exists between these qubits, the switch gate does not introduce an additional constraint. More importantly, if a two-qubit gate involves one qubit in the switching pair but not the other, the temporal dependency created by \switchgate does not cause serialization since the coupler control is distinct from the qubit control, and the zero duration of the switch gate ensures that there is no increase in total execution time.

The switch delay gate \switchdelay would normally increase the circuit duration by the switch response time. However, this delay can be hidden when switching occurs during concurrent two-qubit operations. We implement two optimization strategies to imitate this effect: (a) omit the delay gate when both qubits involved in the switch are simultaneously engaged in two-qubit gates, and (b) omit the delay gate when the temporal position of a neighboring two-qubit gate exceeds the temporal position of the single-qubit predecessor plus the switching duration. These optimizations ensure that control transitions are scheduled within existing idle periods, minimizing the impact on overall circuit execution time and representing a realistic switch scheduling.

\begin{figure}
    \centering
    \includegraphics[width=1\linewidth]{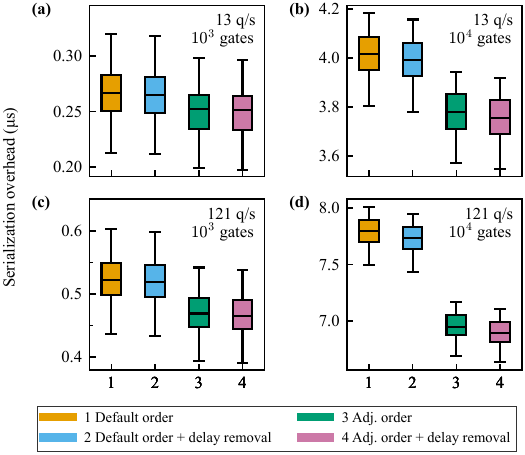}
    \caption{Comparison of optimization strategies for serialization overhead on an 11$\times$11 square grid with 
    (a, c) 1000 and (b, d) 10000 random gates, using 
    (a, b) 13 and (c, d) 121 qubits per switch. Strategies 1 and 2 use default gate ordering, while 3 and 4 prioritize single-qubit gates based on layer distance to the next two-qubit gate. Strategies 2 and 4 employ delay-gate removal when switching occurs during two-qubit gate execution.}
    \label{fig:optimization-impact}
\end{figure}

In \figref{fig:optimization-impact}, we show examples of the impact on serialization overhead of using these strategies for improving the compilation. We compare results from using the default gate order (orange) to either using optimized delay removal (blue), adjusting gate order (green), or doing both (purple). From the plots, we see that both optimizations reduce the serialization overhead, with the major contribution coming from adjusting the gate order.


\section{Strategies for grouping qubits}
\label{app:grouping}

\begin{figure}
    \centering
    \includegraphics[width=1\linewidth]{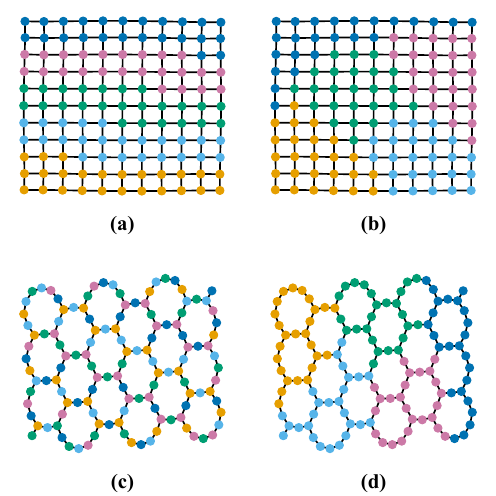}
    \caption{Different strategies for grouping qubits on switches for single-qubit control, illustrated for 26 qubits per switch. 
    (a) Trivial layout for an 11$\times$11 grid. 
    (b) Clustered layout for an 11$\times$11 grid. 
    (c) Dispersed layout for a 127-qubit heavy-hexagon architecture. 
    (d) Clustered layout for a 127-qubit heavy-hexagon architecture.}
    \label{fig:layouts}
\end{figure}

Here, we provide details on and comparisons between the four strategies we have explored for grouping qubits, i.e., to assign groups of $k$ qubits to switches for control of single-qubit gates. In \figref{fig:layouts}, we illustrate three of these strategies: trivial, clustered, and dispersed layouts. In the main text, as discussed in \secref{sec:modeling-serialization-1q}, we are solely using the trivial layout, which is built by assigning qubits to groups following their index. Since qubit indices are typically assigned by row-major counting, such layouts are relatively clustered. The clustered layouts are created using a heuristic balanced clustering algorithm with swap refinements to improve the number of fully-connected clusters and the connectedness of each cluster subgraph over the trivial layouts. The dispersed layouts are based on an algorithm that finds the $d_\text{max}$-distance coloring using a $d$-coloring algorithm and using the maximal $d$ for which such a layout can be found. Finally, the fourth grouping strategy is a random layout, i.e., distributing the qubits randomly among the switches.

\begin{figure}
    \centering
    \includegraphics[width=1\linewidth]{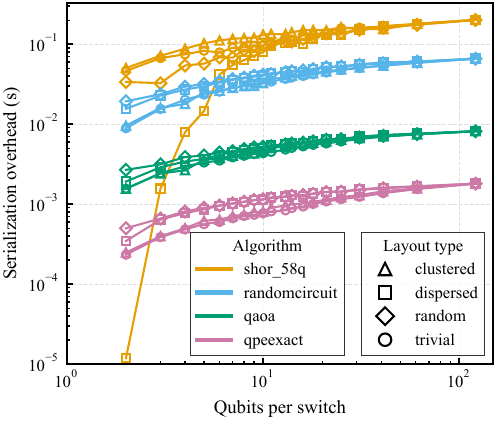}
    \caption{Comparison of the performance of different qubit-to-switch layouts (different markers) on an 11$\times$11 grid. The plot shows serialization overhead for four types of algorithms (different colors). With the notable exception of Shor's algorithm, the clustered and trivial layouts show similarly low overhead, below that of the random and dispersed layouts.}
    \label{fig:layout-overhead-grid121}
\end{figure}

In \figref{fig:layout-overhead-grid121}, we compare the performance of the described layouts through their influence on the serialization overhead, using random circuits and a few selected quantum algorithms from the MQT Bench set. For most algorithms, the overhead is similar for the different alyouts. In general, the influence of the chosen layout decreases for increasing $k$ until the same value is reached for $k=n$, where $n$ is the total number of qubits on the processor.
Among the selected quantum algorithms, only the circuit that implements Shor's algorithm shows large deviations for small $k$. In this case, the dispersed layout adds only \SI{10}{\micro\second} serialization overhead for $k=2$ to the circuit execution while other layouts add up to \SI{30}{\milli\second}, which is related to the structure of the circuits in this algorithm. For the other algorithms, including random circuits, we see that the trivial and clustered layouts tend to yield somewhat lower overhead. This observation is the reason we selected the trivial layout for all other investigations in this article.


\section{Additional results for square-grid qubit layouts}
\label{app:square-grid}

\begin{figure*}
    \centering
    \includegraphics[width=1\linewidth]{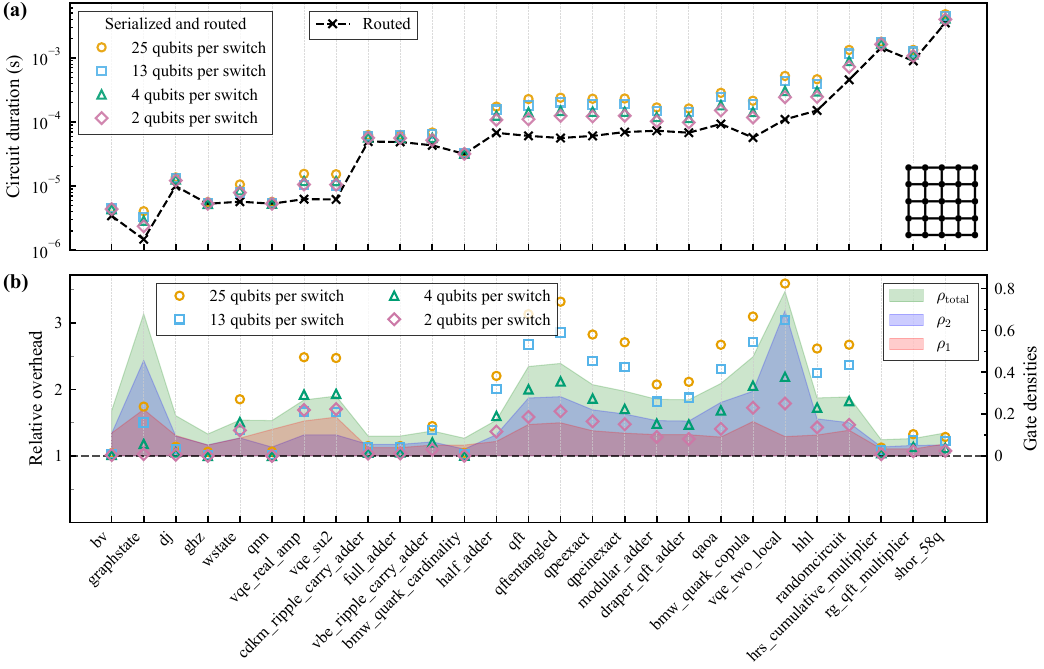}
    \caption{Impact of time multiplexing on execution of quantum algorithms from the MQT Bench set for a 5$\times$5 grid. 
    (a)~Total circuit duration for serialized routed circuits (colored markers) and routed circuits without serialization (black crosses and dashed line) across different quantum algorithms, ordered by increasing gate count. 
    (b)~Relative overhead of serialization compared to the routed circuit duration (markers, left axis), with shaded areas indicating gate densities ($\rho_1$ pink, $\rho_2$ blue, $\rho_{\rm total}$ green; right axis) and different marker colors indicating different numbers of qubits per switch. The data points are medians over 20 transpiler seeds. The full names of the quantum algorithms are listed in \tabref{tab:fit_parameters} in \appref{app:modeling_scaling}.}
    \label{fig:line-mqt-grid25}
\end{figure*}

In this appendix, we present additional results for the 5$\times$5 and 11$\times$11 square-grid qubit layouts, complementing the plots shown in \secref{sec:results}. 

First, in \figref{fig:line-mqt-grid25}, we show the MQT Bench results for the 5$\times$5 grid, in the same style as for the 11$\times$11 grid in \figref{fig:line-mqt-grid121} in the main text. Figure~\figpanelNoPrefix{fig:line-mqt-grid25}{a} shows the total circuit duration for different numbers of qubits per switch for the quantum algorithms in MQT Bench. \figpanel{fig:line-mqt-grid25}{b} shows the relative increase in circuit duration due to serialization (the ratio of serialized to routed circuit duration). The results for the 5$\times$5 grid display the same patterns as those for the 11$\times$11 grid: serialization overhead varies widely between algorithms, but tends to be larger for algorithms with higher gate densities. However, we note that the absolute and relative serialization overhead is smaller for the 5$\times$5 grid than for the 11$\times$11 grid, even for the same number of qubits per switch. We attribute the difference in absolute serialization overhead to the fact that the benchmark circuits are longer for more qubits; for the difference in relative serialization overhead, we note that it is consistent with the pattern seen in the bottom row in \figref{fig:random-grids}, as discussed at the end of \secref{sec:benchmarks-random}.

\begin{figure*}
    \centering
    \includegraphics[width=1\linewidth]{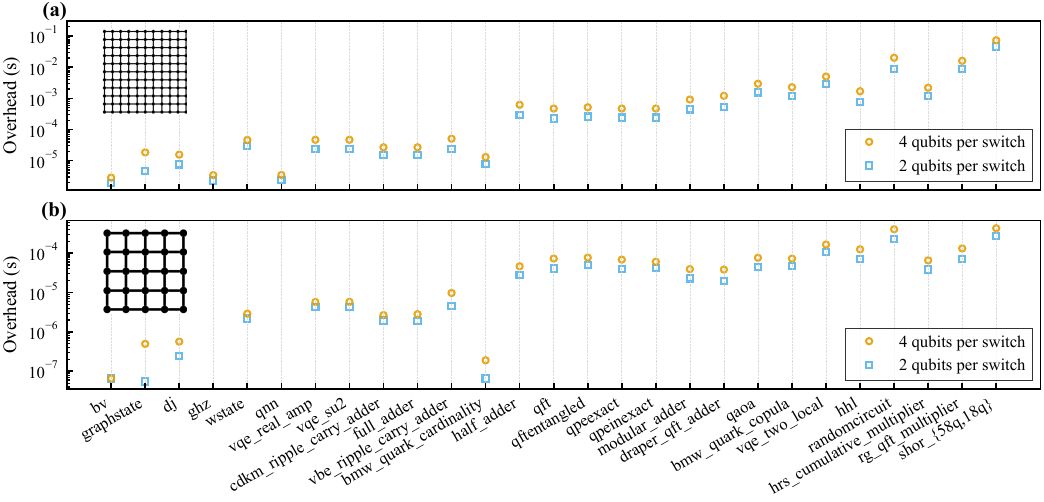}
    \caption{Serialization overhead, in seconds, for the cases of 2 (blue squares) and 4 (orange circles) qubits per switch, for quantum algorithms from the MQT Bench set on a
    (a) 11$\times$11 and
    (b) 5$\times$5 
    square grid of qubits. Exact numerical values for~$k=2$ are shown in Table~\ref{tab:mqt-k2}. The full names of the quantum algorithms are listed in \tabref{tab:fit_parameters} in \appref{app:modeling_scaling}.}
    \label{fig:mqt-detail-grids}
\end{figure*}
Next, we zoom in on the results in \figpanel{fig:line-mqt-grid121}{a} in the main text and plot the duration of only the serialization overhead for the cases of 2 and 4 qubits per switch in \figpanel{fig:mqt-detail-grids}{a}. We provide the same type of plot for the 5$\times$5 grid by zooming in on \figpanel{fig:line-mqt-grid25}{a} to plot the serialization overhead for 2 and 4 qubits per switch in \figpanel{fig:mqt-detail-grids}{b}.
These low numbers of qubits per switch are the most likely to be used in first implementations of time-multiplexed qubit control, which makes extracting the duration of the serialization overhead particularly important here, allowing comparison to state-of-the-art qubit coherence time to judge the feasibility of using switches.

\begin{table}
\centering
\caption{Serialization overhead for different platforms and algorithms with $k=2$ qubits/switch. All values in \si{\milli\second}. The full names of the quantum algorithms are listed in \tabref{tab:fit_parameters} in \appref{app:modeling_scaling}.}
\label{tab:mqt-k2}
\renewcommand{\arraystretch}{1.2}
\renewcommand{\tabcolsep}{0.10cm}
\begin{tabular}{lccc}
\toprule
 Algorithm & 5$\times$5 & 11$\times$11  & 127 Hex \\
\midrule
bmw\_quark\_cardinality & 0.000 & 0.008 & 0.011 \\
bmw\_quark\_copula & 0.047 & 1.166 & 5.299 \\
bv & 0.000 & 0.002 & 0.018 \\
cdkm\_ripple\_carry\_adder & 0.002 & 0.016 & 0.045 \\
dj & 0.000 & 0.008 & 0.024 \\
draper\_qft\_adder & 0.020 & 0.542 & 1.501 \\
full\_adder & 0.002 & 0.016 & 0.045 \\
ghz & 0.000 & 0.002 & 0.002 \\
graphstate & 0.000 & 0.005 & 0.060 \\
half\_adder & 0.028 & 0.286 & 1.726 \\
hhl & 0.072 & 0.767 & 2.173 \\
hrs\_cumulative\_multiplier & 0.038 & 1.169 & 6.996 \\
modular\_adder & 0.023 & 0.443 & 1.531 \\
qaoa & 0.044 & 1.561 & 9.028 \\
qft & 0.041 & 0.227 & 0.908 \\
qftentangled & 0.050 & 0.259 & 1.199 \\
qnn & 0.000 & 0.002 & 0.003 \\
qpeexact & 0.039 & 0.240 & 1.042 \\
qpeinexact & 0.042 & 0.240 & 0.989 \\
randomcircuit & 0.225 & 8.916 & 59.625 \\
rg\_qft\_multiplier & 0.070 & 8.696 & 33.883 \\
shor & 0.275 & 46.323 & 168.742 \\
vbe\_ripple\_carry\_adder & 0.005 & 0.023 & 0.086 \\
vqe\_real\_amp & 0.004 & 0.024 & 0.081 \\
vqe\_su2 & 0.004 & 0.024 & 0.090 \\
vqe\_two\_local & 0.107 & 2.827 & 15.829 \\
wstate & 0.002 & 0.029 & 0.014 \\
\bottomrule
\end{tabular}
\end{table}

\begin{figure}
    \centering
    \includegraphics[width=1\linewidth]{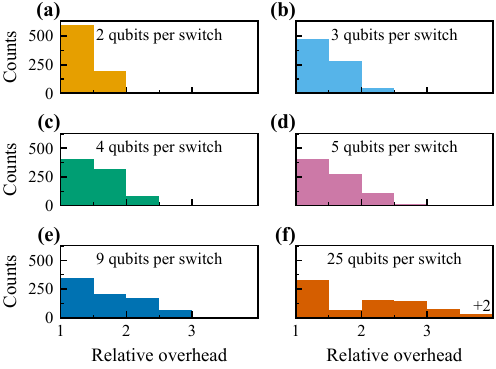}
    \caption{Distribution of time multiplexing overhead across switch sizes on a 5$\times$5 grid, for the quantum algorithms from MQT Bench shown in \figref{fig:line-mqt-grid25}. The histograms show the relative overhead (ratio of serialized to routed circuit duration) for different numbers of qubits per switch. The $+n$ notation to the right in the bottom right histogram indicates $n$ additional counts beyond the histogram limits.}
    \label{fig:hist-mqt-grid25}
\end{figure}

Finally, in the same style as \figref{fig:hist-mqt-grid121} in the main text, we present the data from \figpanel{fig:line-mqt-grid25}{b} in another form in \figref{fig:hist-mqt-grid25}. Here, we display histograms showing the distribution of serialization overhead relative to the routed circuit duration, for six different numbers of qubits per switch, ranging from 2 to 25. These results for the 5$\times$5 grid follow the same pattern as those for the 11$\times$11 grid in \figref{fig:hist-mqt-grid121}: as the number of qubits per switch increases, large serialization overheads gradually become increasingly common.


\section{Results for a heavy-hexagon qubit layout}
\label{app:heavy-hexagon}

In this appendix, we present results for serialization overhead with 127 qubits in a heavy-hexagon layout, corresponding to those shown for 5$\times$5 and 11$\times$11 square-grid layouts in \secref{sec:results} in the main text and in \appref{app:square-grid}. Note that the number of qubits is almost the same for this heavy-hexagon layout as for the 11$\times$11 square-grid layout.

\begin{figure}
    \centering
    \includegraphics[width=1\linewidth]{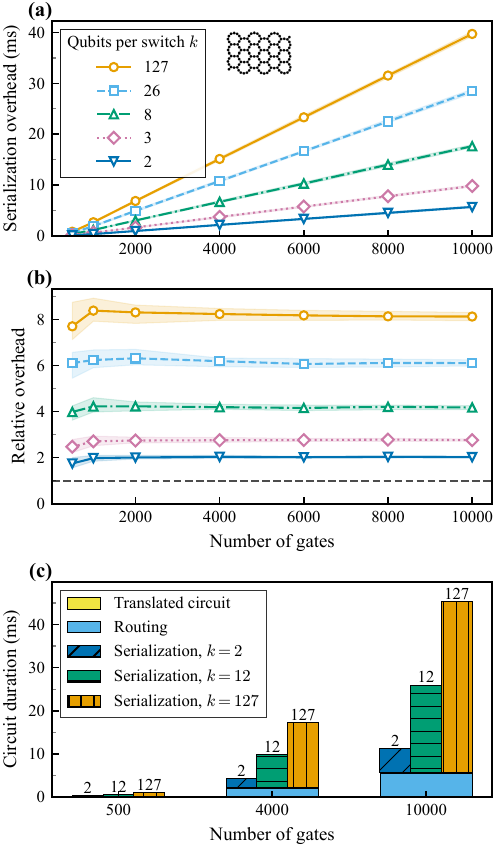}
    \caption{Overhead from time multiplexing single-qubit control for a 127-qubit heavy-hexagon architecture. Each data point displayed is the median over 100 seeds creating random circuits; shaded areas show the interquartile range.  
    (a) Serialization overhead as increase in circuit duration compared to the routed circuit, as a function of the number of gates in random circuits. 
    (b) The same serialization overhead as in~(a), but now shown relative to the duration of the routed circuit, as a function of the number of gates in random circuits. 
    (c) Breakdown of total circuit duration: translated circuit (black), qubit routing overhead (light blue), and serialization overhead (blue, green, yellow), for three different numbers of gates in random circuits.
    The procedure for generating and compiling the circuits is detailed in \secref{sec:time-multiplexing}.}
    \label{fig:line-random-eagle}
\end{figure}

In \figref{fig:line-random-eagle}, we show the overhead due to time multiplexing for random circuits, in the way as we did for square grids in \figref{fig:random-grids} in the main text. We observe the same patterns for the heavy-hexagon layout in \figref{fig:line-random-eagle} as for the square grids in \figref{fig:random-grids}: in \figpanel{fig:line-random-eagle}{a}, we see that the serialization overhead increases linearly with the number of gates in the circuits; in \figpanel{fig:line-random-eagle}{b}, we see that the ratio between the duration of the serialized circuits and the duration of the routed circuit remains constant as a function of the number of gates in the circuits (after some minor variation at the start); in \figpanel{fig:line-random-eagle}{c}, we see that routing itself gives a substantial overhead compared to the uncompiled circuit, but that having more than 2 qubits per switch gives an even larger overhead. Compared to the results for the 11$\times$11 square-grid layout in the right column of \figref{fig:random-grids}, the absolute serialization overhead is longer here for the 127-qubit heavy-hexagon layout, but the relative overhead compared to routing is quite similar. The reason for the increased routing overhead is the sparser connectivity of the heavy-hexagon layout compared to the square-grid layout.

\begin{figure*}
    \centering
    \includegraphics[width=1\linewidth]{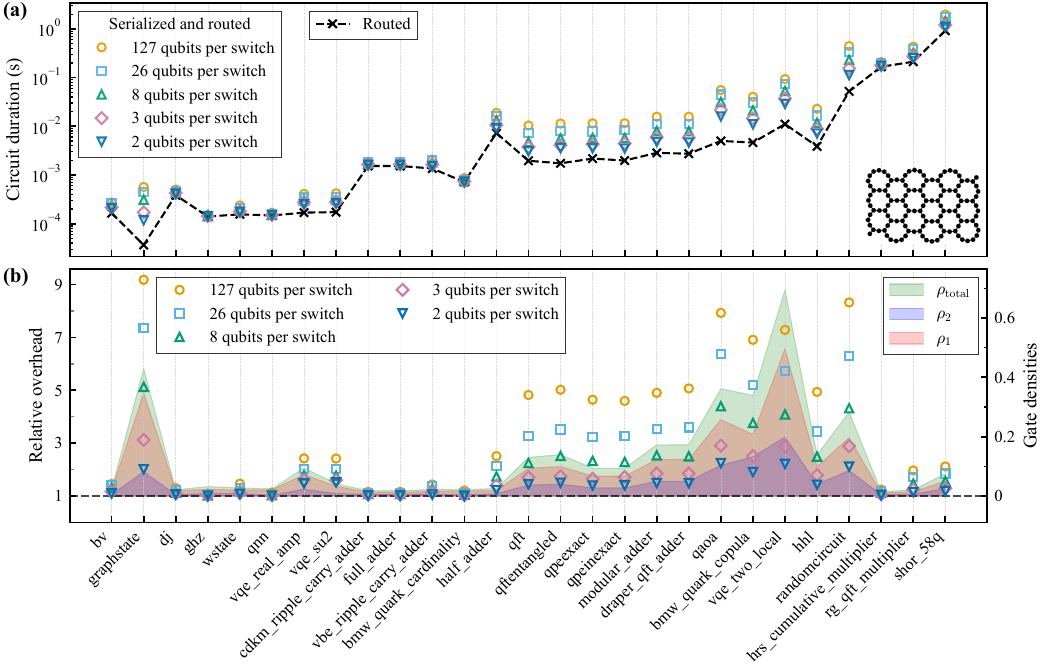}
    \caption{Impact of time multiplexing on execution of quantum algorithms from the MQT Bench set for a 127-qubit heavy-hexagon layout. 
    (a) Total circuit duration for serialized routed circuits (colored markers) and routed circuits without serialization (black crosses and dashed line) across different quantum algorithms, ordered by increasing gate count. 
    (b) Relative overhead of serialization compared to the routed circuit duration ((markers, left axis)), with shaded areas indicating gate densities ($\rho_1$ pink, $\rho_2$ blue, $\rho_{\rm total}$ green; right axis) and different marker colors indicating different numbers of qubits per switch. The data points are medians over 20 transpiler seeds. The full names of the quantum algorithms are listed in \tabref{tab:fit_parameters} in \appref{app:modeling_scaling}.}
    \label{fig:line-mqt-eagle}
\end{figure*}

Next, in \figref{fig:line-mqt-eagle}, we show the MQT Bench results for the heavy-hexagon layout, in the same style as for the 11$\times$11 square grid in \figref{fig:line-mqt-grid121} in the main text and for the 5$\times$5 square grid in \figref{fig:line-mqt-grid25} in \appref{app:square-grid}. Figure~\figpanelNoPrefix{fig:line-mqt-eagle}{a} shows the total circuit duration for different numbers of qubits per switch for the quantum algorithms in MQT Bench, while \figpanel{fig:line-mqt-eagle}{b} shows the serialization overhead for the same cases, relative to the duration of the routed circuit. The results for the heavy-hexagon layout display the same patterns as those for the 5$\times$5 and 11$\times$11 square grids: serialization overhead varies widely between algorithms, but tends to be larger for algorithms with higher gate densities. Comparing to the 11$\times$11 square grid in \figref{fig:line-mqt-grid121}, we note that the absolute serialization overhead is larger for the heavy-hexagon layout, while the relative serialization overhead is more or less the same across algorithms for both cases. Just as in the paragraph above, we attribute the difference in absolute serialization overhead to the increased routing overhead due to the sparser connectivity of the heavy-hexagon layout.

\begin{figure}
    \centering
    \includegraphics[width=1\linewidth]{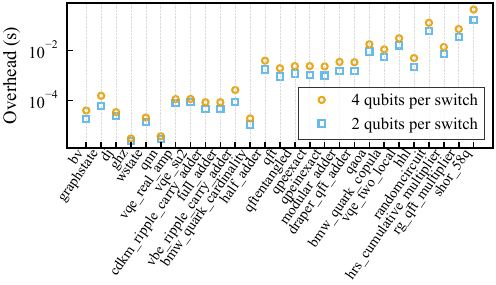}
    \caption{Serialization overhead, in seconds, for the cases of 2 (blue squares) and 4 (orange circles) qubits per switch, for quantum algorithms from the MQT Bench set on a 127-qubit heavy-hexagon architecture. Exact numerical values for $k=2$ are shown in Table~\ref{tab:mqt-k2}. The full names of the quantum algorithms are listed in \tabref{tab:fit_parameters} in \appref{app:modeling_scaling}.}
    \label{fig:mqt-detail-eagle}
\end{figure}

We now zoom in on the results in \figpanel{fig:line-mqt-eagle}{a} and plot the duration of only the serialization overhead for the cases of 2 and 4 qubits per switch in \figref{fig:mqt-detail-eagle}, in the same way as we did for the 11$\times$11 grid in \figpanel{fig:mqt-detail-grids}{a} and for the 5$\times$5 grid in \figpanel{fig:mqt-detail-grids}{b}. We see that the results for the 127-qubit heavy-hexagon layout are quite similar to those for the 11$\times$11 grid, with slightly longer serialization overheads in the heavy-hexagon case.

\begin{figure}
    \centering
    \includegraphics[width=1\linewidth]{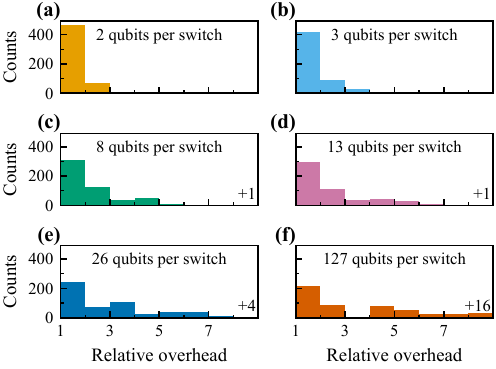}
    \caption{Distribution of time multiplexing overhead across switch sizes on a 127-qubit heavy-hexagon architecture, for the quantum algorithms from MQT Bench shown in \figref{fig:line-mqt-eagle}. The histograms show the relative overhead (ratio of serialized to routed circuit duration) for different numbers of qubits per switch. The $+n$ notation to the right in the some histograms indicates $n$ additional counts beyond the histogram limits.}
    \label{fig:hist-mqt-eagle}
\end{figure}

Finally, in the same style as \figref{fig:hist-mqt-grid121} in the main text for the 11$\times$11 grid and \figref{fig:hist-mqt-grid25} in \appref{app:square-grid} for the 5$\times$5 grid, we present the data from \figpanel{fig:line-mqt-eagle}{b} in another form in \figref{fig:hist-mqt-eagle}. Here, we display histograms showing the distribution of serialization overhead relative to the routed circuit duration, for six different numbers of qubits per switch, ranging from 2 to 127. These results for the heavy-hexagon layout follow the same pattern as those for the 11$\times$11 grid in \figref{fig:hist-mqt-grid121} and the 5$\times$5 grid in \figref{fig:hist-mqt-grid25}: as the number of qubits per switch increases, large serialization overheads gradually become increasingly common. The distributions in the different histograms are very similar to those in the corresponding histograms for the 11$\times$11 grid in \figref{fig:hist-mqt-grid121}.


\section{Modeling the overhead scaling}
\label{app:modeling_scaling}

Here, we present the numerical toy model and the analytical derivation we use to understand the scaling of serialization overhead with the number of qubits per switch $k$, discussed in \secref{sec:scaling}. We also provide further data to complement \figref{fig:prediction-mqt} in that section. 


\subsection{Numerical modeling of overhead scaling}

\begin{figure}
    \centering
    \includegraphics[width=1\linewidth]{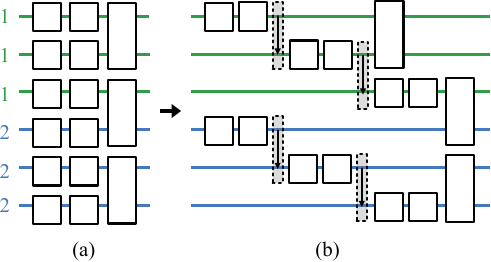}
    \caption{Serialization of a dense circuit. 
    (a) Circuit with unit gate density. 
    (b) The serialized circuit shows extended duration linear in qubits per switch and single-qubit gates. }
    \label{fig:dense-serialization}
\end{figure}

For a dense circuit with many single-qubit gates, as illustrated in \figref{fig:dense-serialization}, not much optimization is possible and we expect a linear scaling in $k$ for the serialization overhead. However, for lower gate densities, more optimization of gate ordering and placement becomes possible. To illustrate the difference between these cases, we put together a simple numerical model.

We model the circuit generation and overhead calculation as follows. First, we set a circuit depth and loop through layers of the circuit. For each qubit in each layer, we add a single-qubit gate with probability $p_1$. Assuming a square grid, we then add a two-qubit gate on nearest-neighbor pairs with probability $p_2$. We normalize units of time such that $t_1=1$.

For each layer in the circuit, we perform two calculations. First, we calculate the ideal runtime, which is unity if there are only parallel single-qubit gates, and $t_2$ if there is a two-qubit gate. Second, we calculate the serialized time per layer. If there is a two-qubit gate, we take the maximum of $t_2$ and the number of single-qubit gates on a switch. If there is no two-qubit gate, the serialized runtime is the number of single-qubit gates. We sum over layers to obtain the serialized runtime over ideal runtime (overhead factor). 

\begin{figure}

    \centering
    \includegraphics[width=\linewidth]{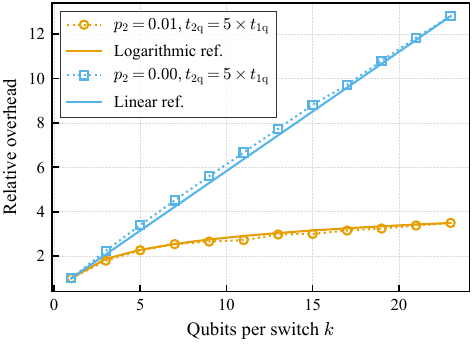}
    \caption{Overhead scaling with number of qubits per switch $k$ for a simple numerical model on a 5$\times$5 grid. The single-qubit gate probability is fixed to $p_1=0.2$, roughly corresponding to the gate density. For a nonzero two-qubit gate probability $p_2 = 0.01$ and two-qubit gate duration $t_2$ larger than single-qubit gate duration $t_1$, the overhead scales as $\log\, k$ (round markers, orange). For $p_2=0$ (squares, blue), the scaling is linear in $k$. The results are averaged over 1000 random samples. The reference curves are scaled to match the data.}
    \label{fig:simple_scaling_model}
\end{figure}

In Fig.~\ref{fig:simple_scaling_model}, we show scaling behaviors for selected parameters using this simple numerical model. We see from the plot that when there are only single-qubit gates, the serialization overhead grows linearly with $k$, but when there are some two-qubit gates present, and the density of single-qubit gates is not so high, the serialization overhead instead grows logarithmically with $k$.


\subsection{Further supporting data}

\begin{figure*}
    \centering
    \includegraphics[width=1\linewidth]{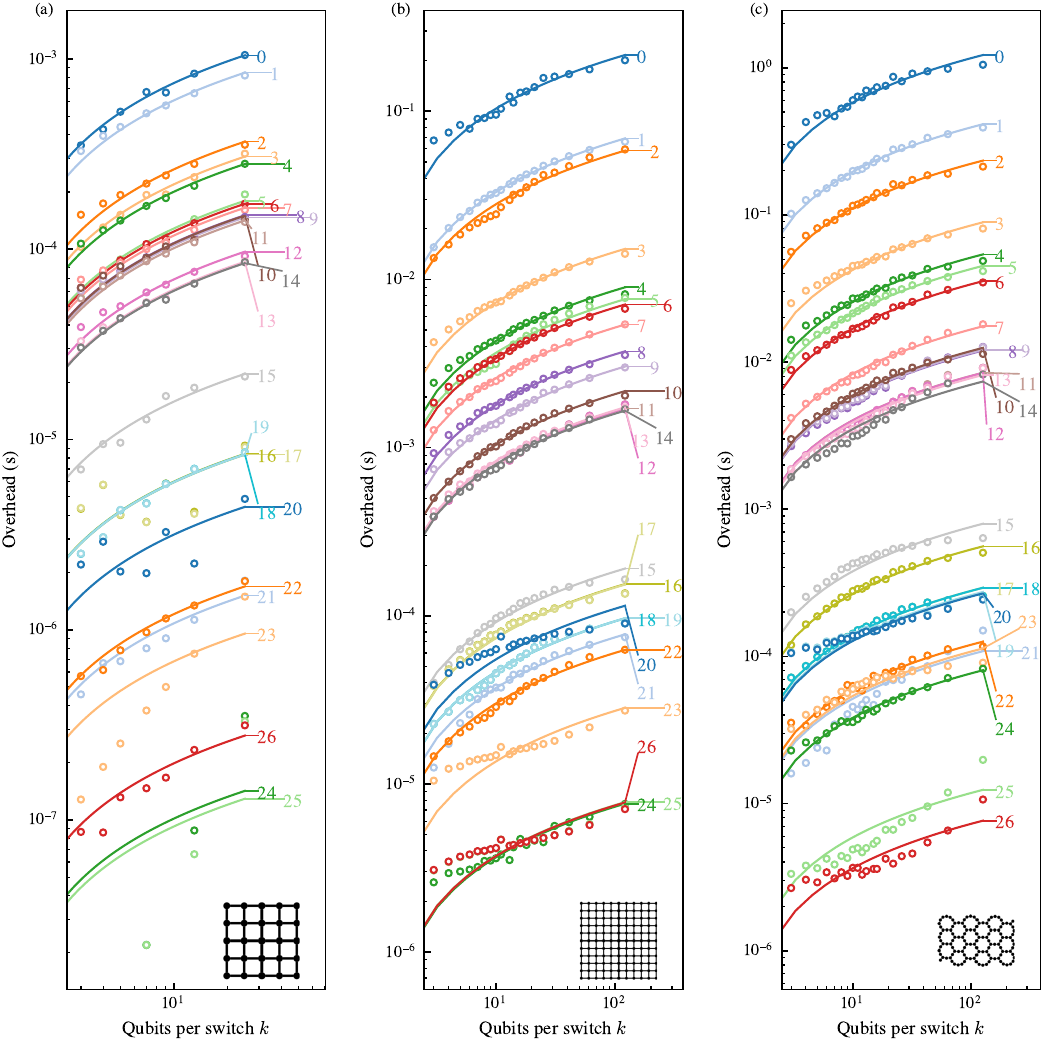}
    \caption{Scaling of serialization overhead for the quantum algorithms from the MQT Bench set on
    (a) a 5$\times$5 grid,
    (b) an 11$\times$11 grid, and
    (c) a 127 qubit heavy-hexagon layout. Numbers label for different quantum algorithms, as listed in \tabref{tab:fit_parameters}. The fit function is $T_\Delta(k) = p N_1^\text{routed}t_\text{1q}\log(k)$. The fit parameters $p$ for all fitted lines (solid curves) are listed in Table \ref{tab:fit_parameters}.}
    \label{fig:fits-mqt-all}
\end{figure*}

In \figref{fig:fits-mqt-all}, we expand on the selection in \figref{fig:prediction-mqt} in the main text, showing the full data for all algorithms from the MQT Bench set, and for all three quantum processor layouts considered in this article, for scaling of serialization overhead with the number of qubits per switch~$k$. From these plots and the fits there (solid curves), we see that the logarithmic scaling with $k$ is prevalent. We provide the fitting parameter for each case, along with the full names for the quantum algorithms considered, in \tabref{tab:fit_parameters}.


\subsection{Queueing-theory-inspired derivation of the logarithmic scaling}

Let $W_1, \dots, W_m$ be independent and identically distributed waiting times for $m$ switches, each with an exponential tail:
\begin{equation}
\Pr(W>x) \sim e^{-\eta x}, \quad x\to\infty.
\end{equation}
where $\eta$ is a constant called the asymptotic decay rate. 

Define the maximum
\begin{equation}
M_m = \max\{W_1, \dots, W_m\}.
\end{equation}
The tail of the maximum is
\begin{align}
\Pr(M_m > x) &= 1 - \Pr(M_m \le x) \\
&= 1 - \Pr(W \le x)^m \\
&= 1 - [1 - \Pr(W>x)]^k.
\end{align}

For large $x$, the tail probability is small, \mbox{$u = \Pr(W>x) \ll 1$}, so we can perform a first-order Taylor expansion:
\begin{equation}
(1-u)^k \approx 1 - k u,
\end{equation}
which implies
\begin{equation}
\Pr(M_k > x) \approx k \, \Pr(W>x) \approx k \, e^{-\eta x}.
\end{equation}

To find a maximum, set $\Pr(M_k > x) \sim O(1)$:
\begin{equation}
k e^{-\eta x} \sim 1 \quad \Rightarrow \quad x \sim \frac{1}{\eta} \log k.
\end{equation}
Thus, the maximum waiting time (serialization overhead) grows logarithmically with the number of qubits per switch $k$.

\begin{table*}
\centering
\caption{Full names of the quantum algorithms in the MQT Bench set used in this article, along with the indices used to label them in \figref{fig:fits-mqt-all} and the values for the fitting parameter $p$ for the fits in that figure.}
\label{tab:fit_parameters}
\renewcommand{\arraystretch}{1.2}
\renewcommand{\tabcolsep}{0.30cm}
\begin{tabular}{cccccc}
\hline\hline
Index & Label & Quantum algorithm & 5$\times$5 grid & 11$\times$11 grid & Heavy hexagon \\
\midrule
0 & shor\_58q & Shor's factorization, 58 qubits & 0.005(0) & 0.643(10) & 3.587(64) \\
1 & randomcircuit & Random circuit (2$n$ depth) & 0.019(0) & 1.052(5) & 6.230(46) \\
2 & vqe\_two\_local & VQE Two-local ansatz & 0.050(2) & 1.389(15) & 8.049(109) \\
3 & rg\_qft\_multiplier & Ruiz--Garcia QFT multiplier & 0.003(0) & 0.337(4) & 1.337(15) \\
4 & hhl & HHL linear system solver & 0.080(1) & 1.034(5) & 3.318(30) \\
5 & hrs\_cumulative\_multiplier & HRS cumulative multiplier & 0.007(0) & 0.197(3) & 1.152(10) \\
6 & qaoa & QAOA optimization & 0.038(0) & 1.308(15) & 7.764(93) \\
7 & qftentangled & QFT on GHZ State & 0.133(4) & 0.940(5) & 4.534(83) \\
8 & qft & Quantum Fourier transform & 0.130(3) & 0.960(6) & 4.222(96) \\
9 & qpeexact & QPE exact phase) & 0.118(3) & 0.920(10) & 4.356(89) \\
10 & bmw\_quark\_copula & QUARK copula model & 0.043(2) & 1.356(11) & 6.705(45) \\
11 & qpeinexact & QPE (inexact phase) & 0.114(2) & 0.920(10) & 4.500(73) \\
12 & half\_adder & Half adder & 0.054(1) & 0.811(6) & 4.618(46) \\
13 & modular\_adder & Modular adder & 0.042(1) & 0.994(5) & 3.928(48) \\
14 & draper\_qft\_adder & Draper QFT adder & 0.037(0) & 1.108(8) & 3.514(51) \\
15 & vbe\_ripple\_carry\_adder & VBE ripple carry adder & 0.087(2) & 0.503(8) & 2.077(50) \\
16 & vqe\_real\_amp & VQE real amplitudes & 0.051(6) & 0.634(9) & 1.090(33) \\
17 & vqe\_su2 & VQE efficient SU(2) & 0.047(6) & 0.578(9) & 1.021(32) \\
18 & cdkm\_ripple\_carry\_adder & CDKM ripple carry adder & 0.034(1) & 0.265(3) & 0.795(10) \\
19 & full\_adder & Full adder & 0.034(1) & 0.265(3) & 0.795(10) \\
20 & wstate & W state preparation & 0.040(5) & 0.700(27) & 0.486(6) \\
21 & graphstate & Graph state preparation & 0.030(2) & 1.043(14) & 7.447(80) \\
22 & dj & Deutsch-Jozsa algorithm & 0.018(0) & 0.441(4) & 0.877(11) \\
23 & bmw\_quark\_cardinality & QUARK cardinality model & 0.004(1) & 0.080(3) & 0.302(14) \\
24 & qnn & Quantum neural network & 0.002(1) & 0.066(3) & 0.105(6) \\
25 & ghz & GHZ state preparation & 0.002(1) & 0.097(4) & 0.094(4) \\
26 & bv & Bernstein--Vazirani algorithm & 0.010(0) & 0.176(4) & 2.591(73) \\
\hline\hline
\end{tabular}
\end{table*}


\clearpage

\bibliography{references}

\end{document}